\documentclass[preprint2]{aastex}
\usepackage{natbib}
\usepackage{lscape}
\usepackage{wasysym}
\usepackage{txfonts}
\usepackage{graphicx}
\usepackage{hyperref}
\bibpunct{(}{)}{;}{a}{}{,}
\usepackage{longtable}
\usepackage{hyperref}

\usepackage{afterpage}

\usepackage[usenames,dvipsnames]{color}
\definecolor{mygreen}{rgb}{0,0.5,0}

\def\ms{\hbox{\,m\,s$^{-1}$\,}}         
\def\cms{\hbox{\,cm\,s$^{-1}$\,}}       
\def\m2s2{\hbox{\,m$^{2}$\,s$^{-2}$}\,} 
\def\kms{\hbox{\,km\,s$^{-1}$}\,}       

\newcommand{\titleast}{\ast}
\newcommand{\titlestar}{\star}

\shorttitle{Characterization of a spurious one-year signal in HARPS data}
\shortauthors{X. Dumusque}

\begin{document}

\title{Characterization of a spurious one-year signal in HARPS data\altaffilmark{\titleast}}

\author{Xavier Dumusque\altaffilmark{1}\altaffilmark{\titlestar}
		, Francesco Pepe\altaffilmark{2}
		, Christophe Lovis\altaffilmark{2}
		, David W. Latham\altaffilmark{1}}

\altaffiltext{1}{Harvard-Smithsonian Center for Astrophysics, 60 Garden Street, Cambridge, Massachusetts 02138, USA}
\altaffiltext{2}{Observatoire Astronomique de l'Universit\'e de Gen\`eve, 51 Chemin des Mailettes, 1290 Sauverny, Suisse}

\altaffiltext{$\star$}
{xdumusque@cfa.harvard.edu}

\altaffiltext{$\ast$}
{Based on observations made with the HARPS instrument on the ESO 3.6-m telescope at La Silla Observatory (Chile).}

\begin{abstract}
The HARPS spectrograph is showing an extreme stability close to the \ms level over more than ten years of data. However the radial velocities of some stars are contaminated by a spurious one-year signal with an amplitude that can be as high as a few \ms. This signal is in opposition of phase with the revolution of Earth around the Sun, and can be explained by the deformation of spectral lines crossing block stitchings of the CCD when the spectrum of an observed star is alternatively blue- and red-shifted due to the motion of Earth around the Sun. {\bf This annual perturbation can be supress by either removing those affected spectral lines from the correlation mask used by the cross correlation technique to derive precise radial velocities, or by simply fitting a yearly sinusoid to the RV data. This is mandatory if we want to detect long-period low-amplitude signals in the HARPS radial velocities of quiet solar-type stars.}
\end{abstract}


\section{Introduction} \label{sect:1}

HARPS is a fiber-fed, cross-dispersed echelle spectrograph. The instrument is fed by two fibers, one for the target and the other for a reference lamp or sky. The spectrograph re-images the two fibers on a mosaic of two 2k4 CCDs, where two echelle spectra of 72 orders are formed. The resolution of the spectrograph is R$=$115,000 with a resolution element oversampling of 3.2 CCD pixels \citep[for more information, see][]{Mayor-2003,Pepe-2002}.

{\bf To prevent any large drift in radial velocity (RV), HARPS has been designed to reach an extreme long-term stability of 0.01 mbar in pressure and 0.01 K in temperature. The tiny RV instrumental drifts related to residual pressure and temperature changes are measured by a Thorium-Argon reference lamp \citep[recently upgraded to a Fabry-Perot interferometer calibration source,][]{Wildi-2011}.}

Due to those exceptional characteristics, HARPS, {\bf and now its copy HARPS-North \citep[][]{Cosentino-2012},} has been shown to be the most precise spectrographs to search for extra-solar planets. As an example, the HARPS RV of the magnetically quiet stars $\tau$\,Ceti (\object{HD10700}) and \object{HD85512} show a standard deviation of 0.92\ms and 1.05\ms over more than six years of data, respectively (see Table 2 in \citet{Pepe-2011} for more examples).

The precision of the HARPS spectrograph might be even below the \ms level, as we know that RV measurements are contaminated by different kinds of stellar signals that induce variations that can be as high as a few \ms. To our current knowledge, these stellar signals can be differentiated in four categories: 
\begin{itemize}
\item stellar oscillations produced by pressure waves propagating in the convective zone of the star \citep[][]{Dumusque-2011a,Arentoft-2008,Kjeldsen-2005}
\item stellar granulation that is the result of convection at the stellar surface \citep[][]{Dumusque-2011a}
\item short-term stellar activity induced by stellar rotation in the presence of stellar active regions \citep[][]{Dumusque-2014a,Boisse-2012b,Meunier-2010a,Saar-1997b}
\item long-term stellar activity induced by stellar magnetic cycles \citep[][]{Meunier-2013,Dumusque-2011c}
\end{itemize}

In this paper, we present a new type of perturbing signal that is not induced by the stars themselves, but by the design of the CCD detector. In Section \ref{sect:2} we highlight the presence of this signal in the RVs of several stars observed with HARPS. We then demonstrate the origin of this signal in Section \ref{sect:3} and present a solution to correct for this signal in Section \ref{sect:4}.  Section \ref{sect:5} concludes the paper.

\section{A one-year signal present in HARPS RV data} \label{sect:2}

Several stars that have been intensively observed with HARPS during several years show a low-amplitude signal at a period of one year. This is for example the case of \object{HD128621} ($\alpha$\,Cen\,B), \object{HD1461}, \object{HD154088}, or \object{HD31527}. Note that other stars are also affected by this annual effect.

\subsection{HD128621} \label{sect:2-0}

\begin{figure}
\begin{center}
\includegraphics[width=8cm]{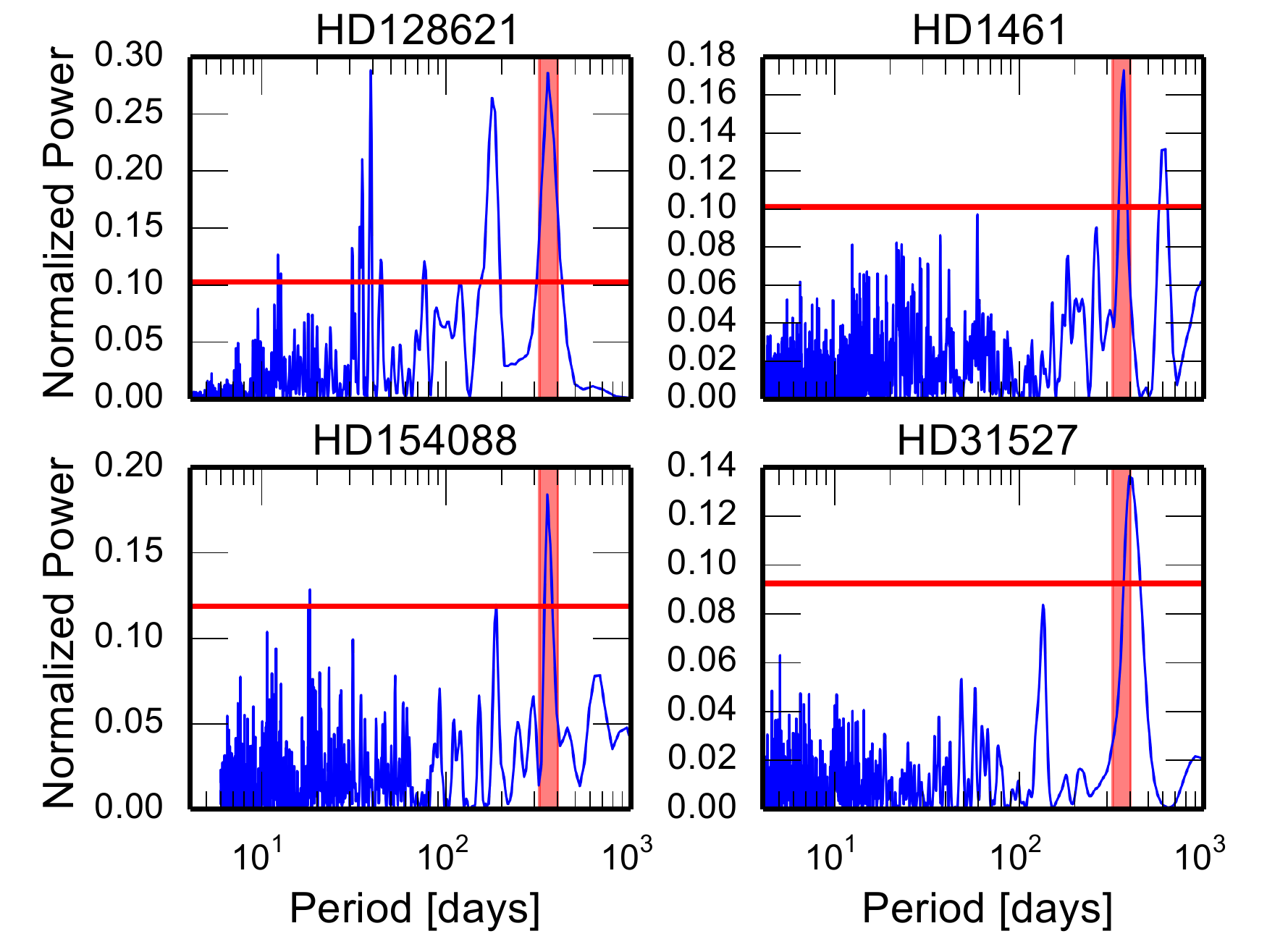}
\caption{{\bf Generalized Lomb Scargle (GLS) periodograms \citep[][]{Zechmeister-2009}} of the RV residuals of our four stars after removing their planetary signals and their magnetic cycle effect (see text in Sec \ref{sect:2} for more details about the fitted models). The red horizontal lines represent a false alarm probability level of 10\%. For each star the one-year signal is significant and is highlighted by the red regions.}
\label{fig:2-0}
\end{center}
\end{figure}

In \citet{Dumusque-2012}, the authors present a detailed analysis of the RV of HD128621 and detect the presence of an Earth-mass planet orbiting at a short period of 3.24 days. In addition to this planetary signal, a detection of the binary RV drift induced by HD128620 ($\alpha$\,Cen\,A), a RV variation induced by active regions, and a long-term drift due to the stellar magnetic cycle are reported. As explained in \citet{Lindegren-2003} and shown in \citet{Dumusque-2011c} and \citet{Meunier-2013}, a stellar magnetic cycle induces a correlated effect in RV because active regions strongly suppress stellar convection. By fitting a 3rd order polynomial to the raw RVs between 2008 and 2011 to remove simultaneously the binary and the magnetic cycle effect, a yearly signal is detected in the residuals (Figure \ref{fig:2-0} top left panel). This one-year signal cannot be detected in the \citet[][]{Dumusque-2012} published RV data, because this effect was discovered at the time, and was corrected by using the method described in Sec. \ref{sect:4}.

\begin{figure}
\begin{center}
\includegraphics[width=8cm]{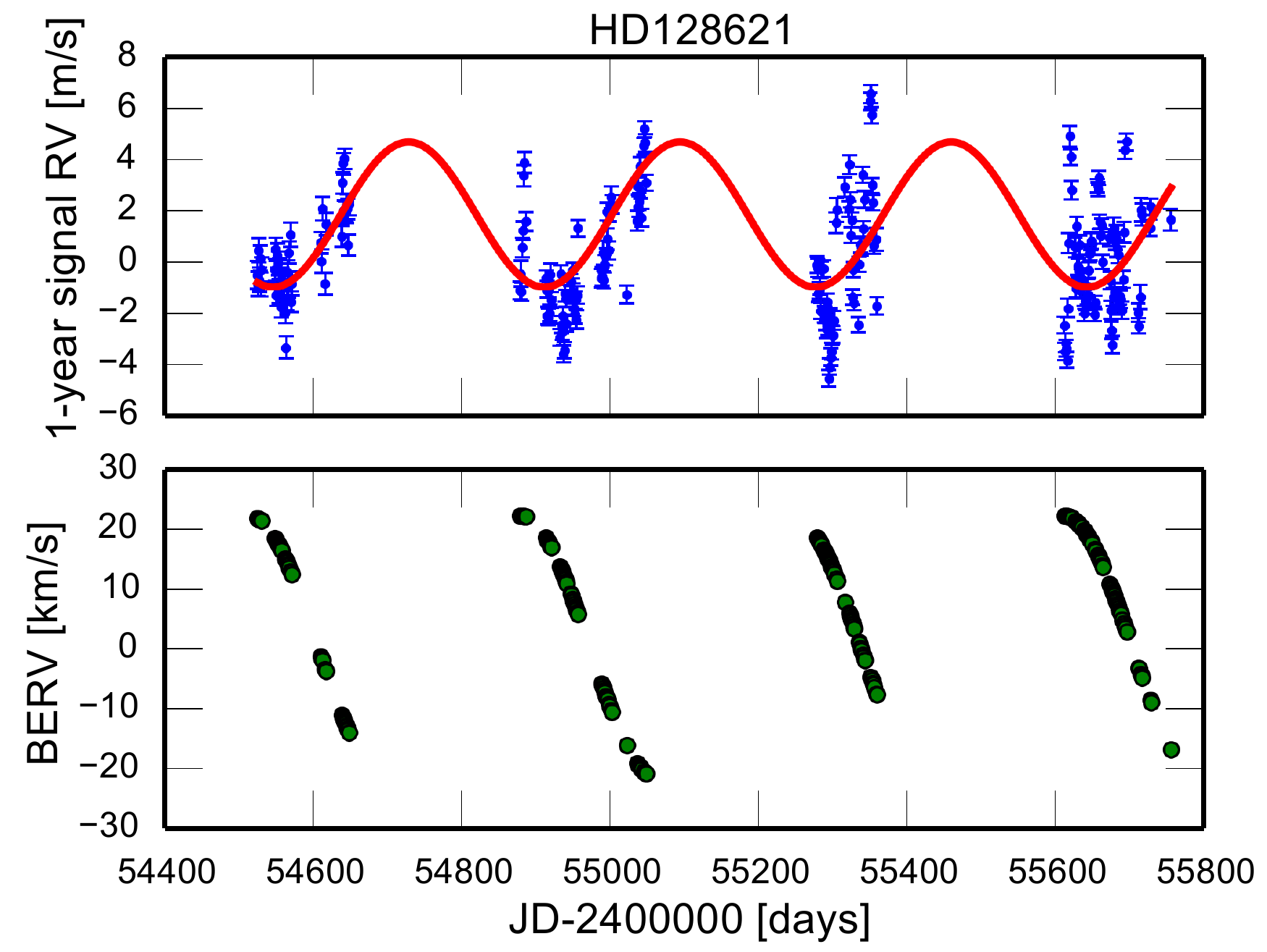}
\caption{\emph{Top: }{\bf RV residuals of HD128621 in blue after fitting a 3rd order polynomial. The red curve corresponds to the fitted one-year signal}. \emph{Bottom: }Barycentric Earth radial velocity (BERV) of HD128621, which is the velocity of the Earth in the direction of HD128621.}
\label{fig:2-1}
\end{center}
\end{figure}

The yearly signal is extracted from the data by fitting a sinusoid with a 365.25-day period, in addition to the 3rd order polynomial. This yearly signal that has an amplitude of 2.83 $\pm$ 0.47\ms is shown in the top panel of Figure \ref{fig:2-1}. The bottom panel of the same figure shows the barycentric Earth radial velocity (BERV), i.e. the velocity of Earth in the direction of HD128621. The stellar velocity is the {\bf relativistic} sum of the raw RV as measured by the spectrograph and the BERV. The BERV varies with a period of one year because Earth orbits the Sun, and its maximum amplitude depends on the position of the star in the sky, but can be as large as $\pm$30\kms. This velocity variation can be translated to a shift of the spectrum on the HARPS CCD of $\pm$37 pixels\footnote{one pixel on the HARPS CCD represent 820\ms}. {\bf Although the strongest signal in the BERV is the one with the one-year period, other signals with different periods are relevant for precise Doppler work at the \ms precision. Examples of these other signals are the rotation of the Earth around its axis, the revolution around the Sun of the solar system planets or relativistic effects \citep[see e.g.][for a detailed review]{Wright-2014}. In this study, we focus on the yearly signal because it is the only one that seem to be detected in the data.}


The fact that an anti-correlation is observed between the BERV for HD128621 and the one-year signal detected in the RVs (Figure \ref{fig:2-2}, top left panel) indicates that the two yearly signals are in opposition of phase. The BERV is the velocity of the Earth in the direction of the star, and the RV is the measure of the stellar velocity in the direction of Earth. Therefore this opposition in phase {\bf suggests} that the one year signal detected in the RVs is induced by the orbit of Earth around the Sun and is not corresponding to an extrasolar planet.

\subsection{HD1461} \label{sect:2-1}

Diaz et al. 2015 (in prep.) analyzed the HARPS RV data of HD1461 and found the presence of two planets orbiting at 5.77 and 13.50 days with 0.05 and 0.09 of eccentricity, respectively. In addition, the star also shows a long-term magnetic cycle in the variation of its calcium activity index. The effect of the magnetic cycle is seen in the RVs of HD1461, and Diaz et al. 2015 (in prep.) fit it using a Keplerian with a period of 3508 days and an eccentricity of 0.10. 

After fitting a three Keplerian model to the RVs of HD1461 with parameters fixed to these previous values, the RV residuals show a signal at a one-year period (Figure \ref{fig:2-0} top right panel). This signal has an amplitude of 1.31 $\pm$ 0.20\ms when a model with three Keplerians plus a one year sinusoid is fitted to the data.

As for HD128621, the anti-correlation between the BERV for HD1461 and the one year signal seen in the RVs of the star shows that the two signals are in opposition of phase (Figure \ref{fig:2-2} top right panel). Therefore the revolution of Earth around the Sun {\bf seems to be} detected in the RVs of HD1461.

\subsection{HD154088} \label{sect:2-2}

\citet{Mayor-2011} announced the detection of a super-Earth orbiting around HD154088. After fitting a Keplerian model with the planet parameters reported in \citet{Mayor-2011} and a 2nd order polynomial drift to account for the magnetic cycle, the RV residuals show a significant signal at a one-year period (Figure \ref{fig:2-0} bottom left panel). This signal has an amplitude of 0.76 $\pm$ 0.16\ms when a model with one Keplerian, a 2nd order polynomial, plus a one year sinusoid is fitted to the data.

As for HD128621 and HD1461, the anti-correlation between the BERV for HD154088 and the one year signal seen in the RVs {\bf suggests} that the signal of Earth orbiting the Sun is detected (Figure \ref{fig:2-2} bottom left panel).

\subsection{HD31527} \label{sect:2-3}

A planetary system of three planets orbiting HD31527 has been announced by \citet{Mayor-2011}. The activity index variation for the star is very low, with an average log(R'$_{HK}$) level of -4.96. We therefore do not expect any impact of activity on the RVs. When fitting the three planets with the parameters reported in \citet{Mayor-2011}, a signal with a period close to one year is seen in the RV residuals (Figure \ref{fig:2-0} bottom right panel). This signal has an amplitude of 0.59 $\pm$ 0.15\ms when a model with three Keplerians plus a one year sinusoid is fitted to the data.

\begin{figure*}[!th]
\begin{center}
\includegraphics[width=16cm]{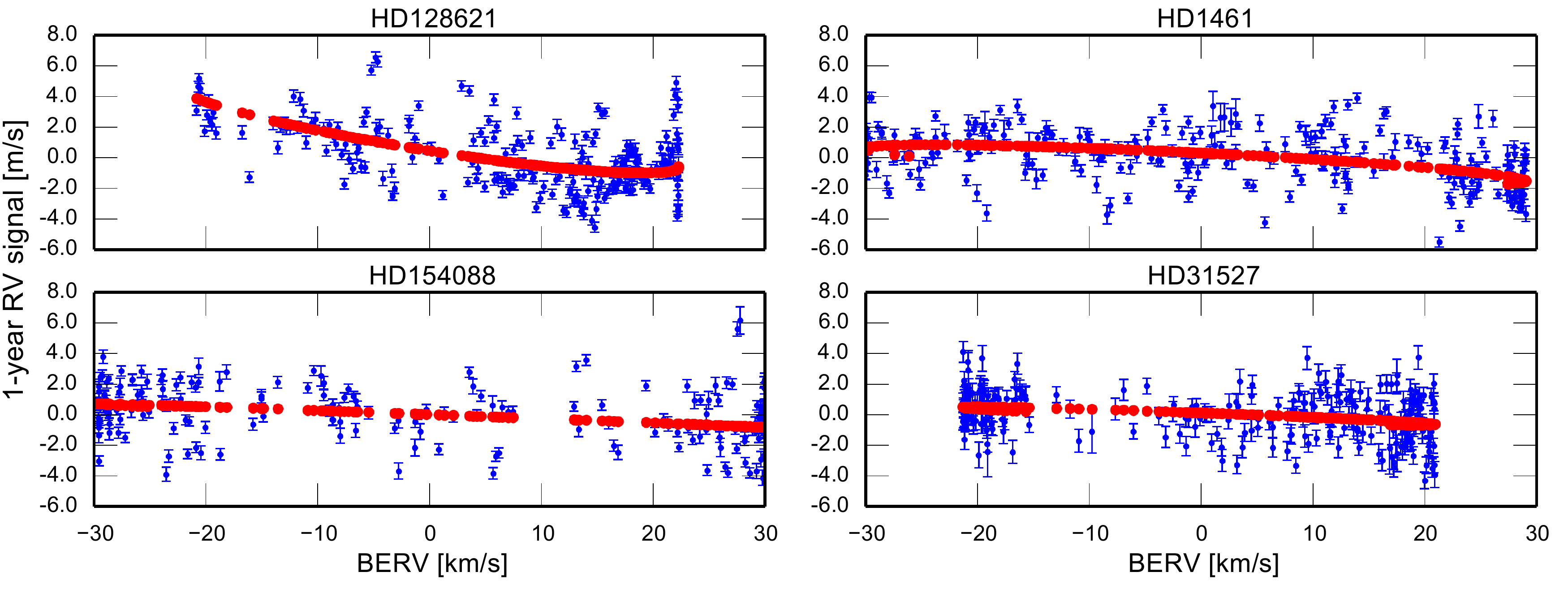}
\caption{Correlations between the RV induced by the one year signal and the barycentric Earth radial velocity (BERV) for our four stars {\bf in red. The blue points represent the underlying data with error bars.}}
\label{fig:2-2}
\end{center}
\end{figure*}

As for the other stars studied in this paper, the opposition in phase between the one-year signal detected in the RV residuals and the BERV for HD31527 {\bf suggests} that the signal of Earth is detected in the stellar RVs (Figure \ref{fig:2-2} bottom right panel).

\section{Origin of the one-year signal} \label{sect:3}

We show in the preceding section that HARPS RV data of several stars that have been intensively observed during several years are affected by a one-year signal that is in opposition of phase with the yearly variation of the BERV of the target. {\bf The same opposition in phase observed in the RV of several stars cannot be a coincidence. It suggests that the signal that we are detecting is due to the Earth orbiting the Sun. Several different effects could be responsible for this annual signal:
\begin{itemize}
\item imprecision in the code to calculate the BERV \citep[][]{Wright-2014,Kopeikin-1999},
\item incorrect stellar coordinates,
\item telluric absorption line contamination,
\item incorrect observatory coordinates,
\item incorrect reduction of observing times,
\item uncorrected systematics in seasonal variations at the telescope,
\end{itemize}
and others. It is difficult to rule out an effect by studying a group of stars, because each star will have a different BERV. However, this is not the goal here and we will demonstrate that the yearly signal is, in the presented example, mostly created by imperfections of the CCD that induce a deformation of spectral lines when the stellar spectrum is red- and blue-shifted on the CCD due to Earth orbiting the Sun.}

\subsection{Measuring the radial velocity of individual spectral lines}  \label{sect:3-1}

When Earth is revolving around the Sun, the BERV of a target is changing as a function of time, and this translates into a shift of the recorded spectrum on the CCD due to the Doppler effect. If a spectral line passes over an imperfection of the CCD, this spectral line can undergo a deformation that will introduce a RV variation correlated or anti-correlated with the BERV.
\begin{figure}[!t]
\begin{center}
\includegraphics[width=8cm]{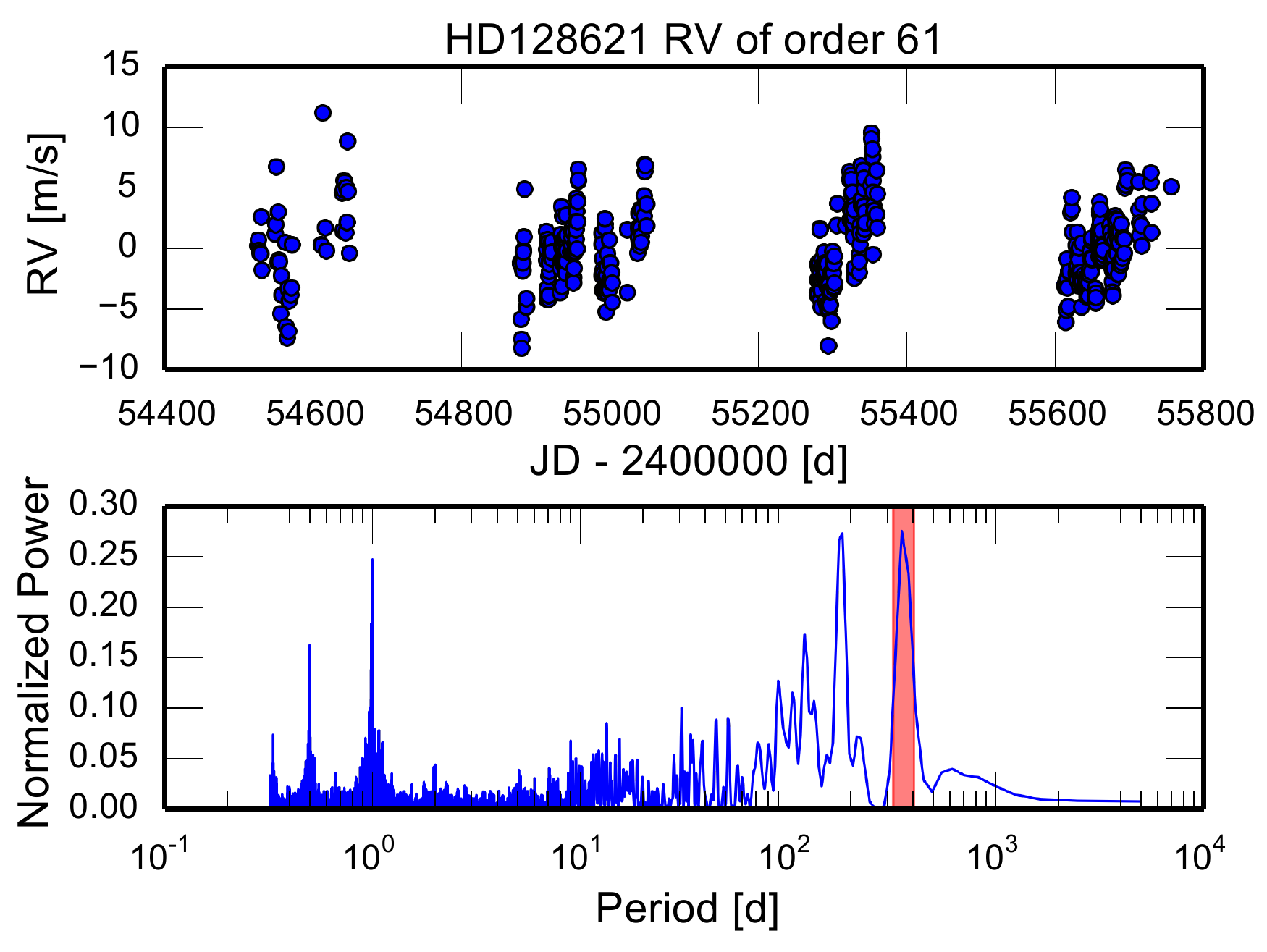}
\caption{RV residuals as a function of time for HD128621 derived from the HARPS spectral order 61, after fitting a 3rd order polynomial to the data. The one-year signal is significant in the corresponding GLS periodogram (red region).}
\label{fig:3-0}
\end{center}
\end{figure}

To test this hypothesis, we first looked at the RV of each spectral order individually, as some spectral orders might be more affected than others. In Figure \ref{fig:3-0}, we can see the RVs derived using only the spectral order 61. This spectral order spans a wavelength between 6145 and 6215 \AA. {\bf Despite the absence of relevant telluric lines, the RV of this spectral order is nevertheless} strongly affected by a one-year signal.

\begin{figure*}
\begin{center}
\includegraphics[width=8cm]{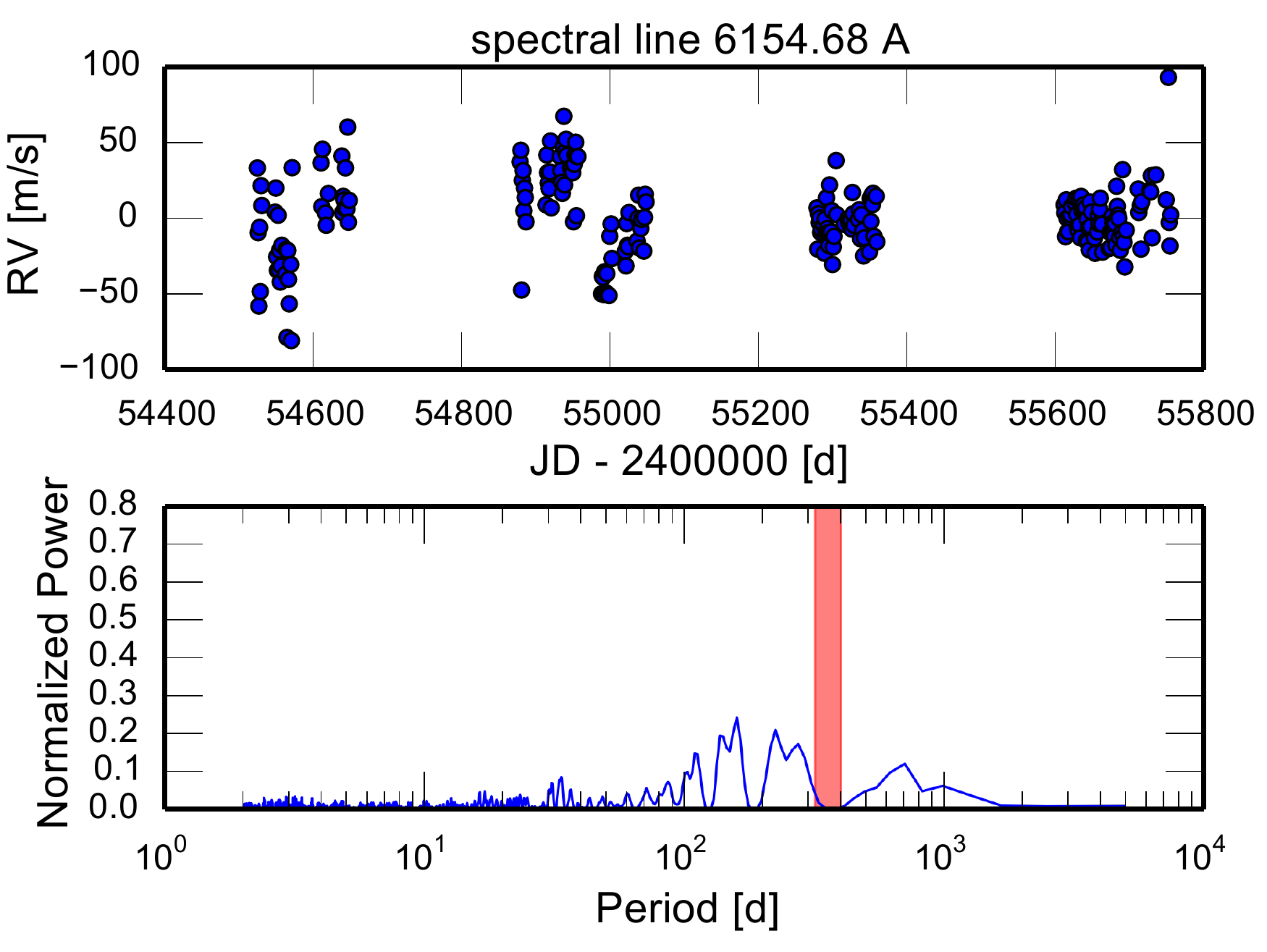}
\includegraphics[width=8cm]{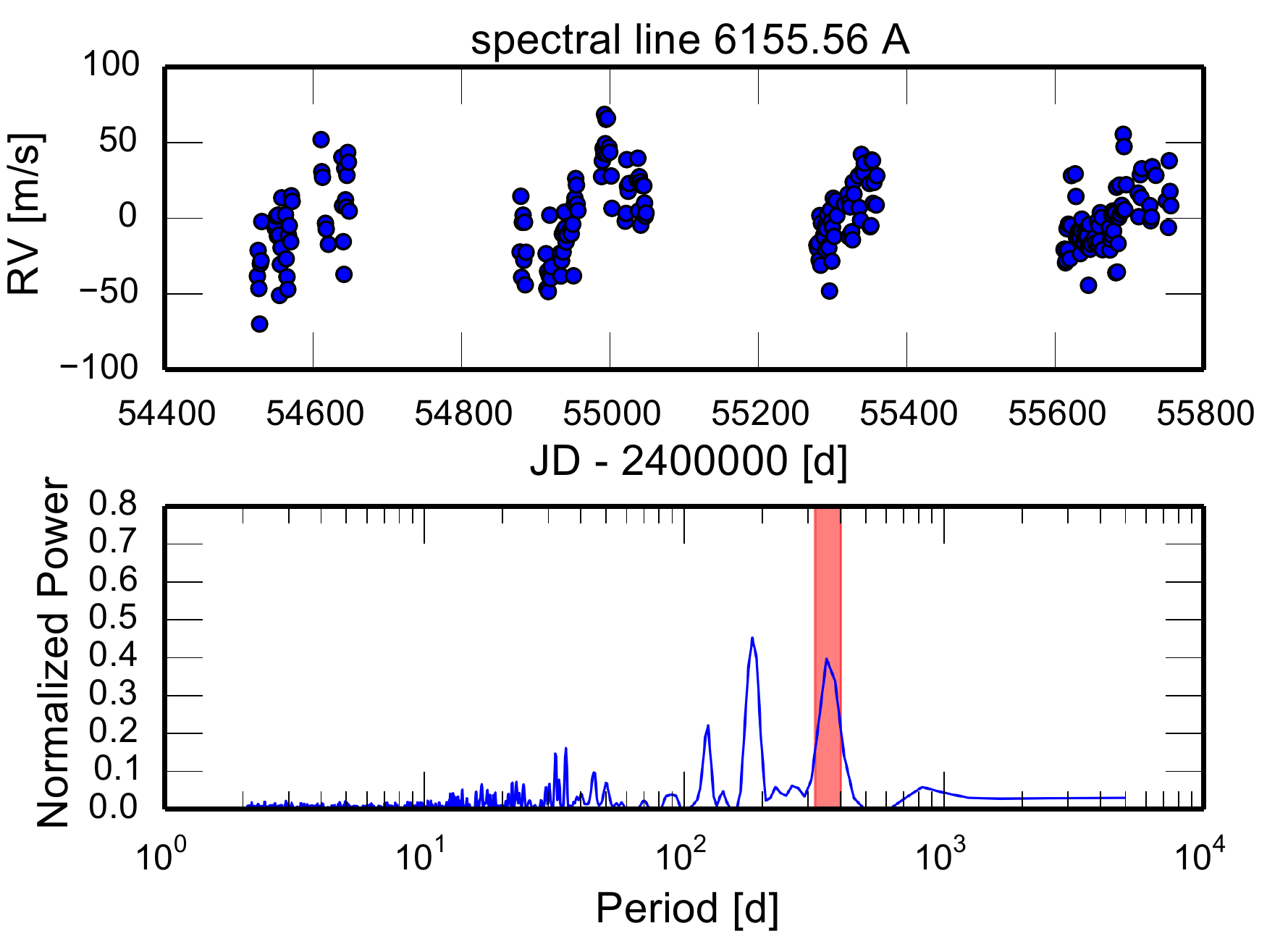}
\includegraphics[width=8cm]{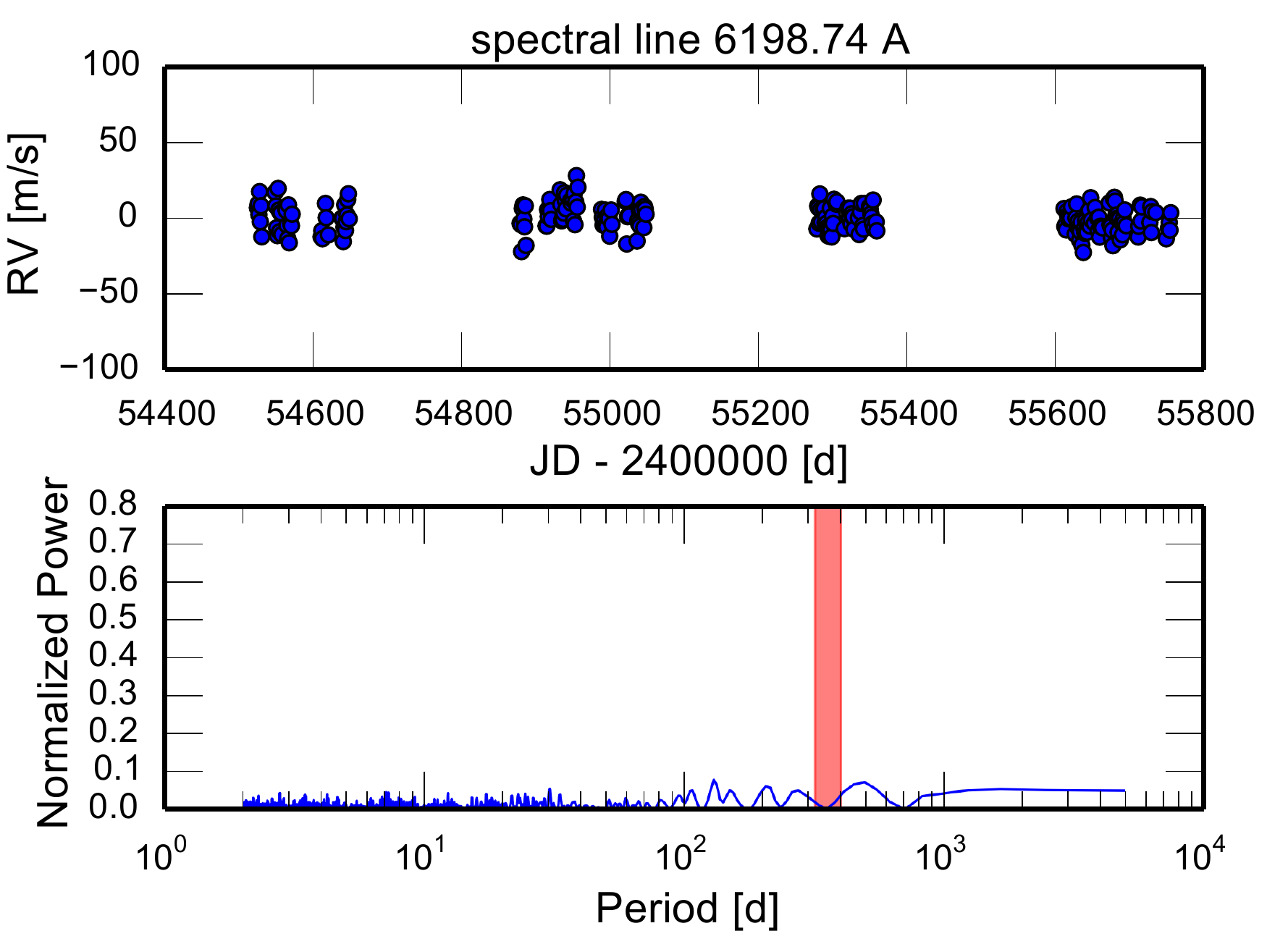}
\includegraphics[width=8cm]{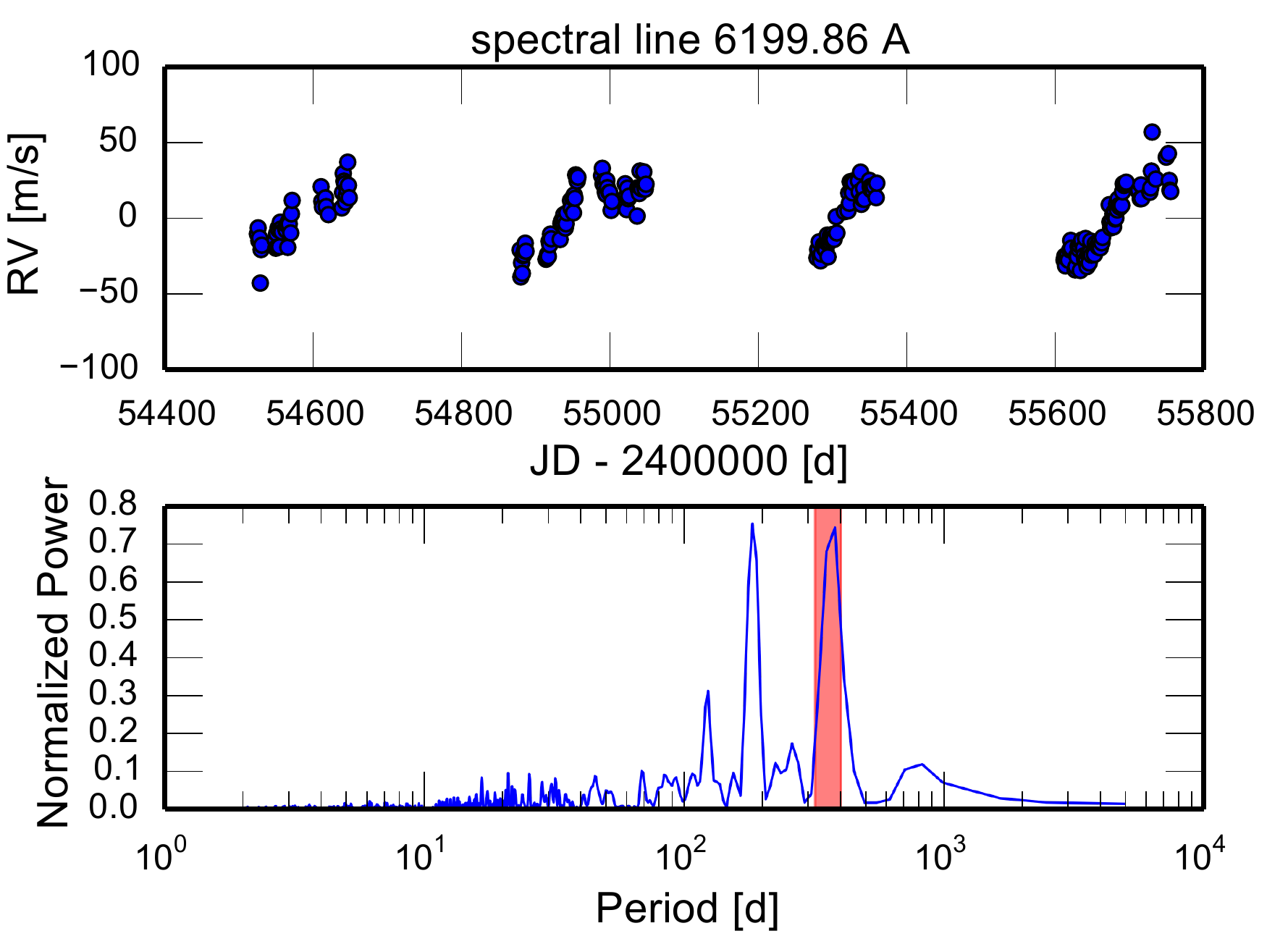}
\caption{RV residuals as a function of time for HD128621 derived from different spectral lines present in the HARPS spectral order 61, after fitting a third order polynomial to the RV of each line. The corresponding GLS periodograms are also shown and highlight in red the {\bf presence of the one-year signal for spectral lines at 6155.56 and 6199.86 \AA\,, and the absence of the signal for the nearby lines at 6154.68 and 6198.74 \AA.}}
\label{fig:3-1}
\end{center}
\end{figure*}

\begin{figure*}
\begin{center}
\includegraphics[width=8cm]{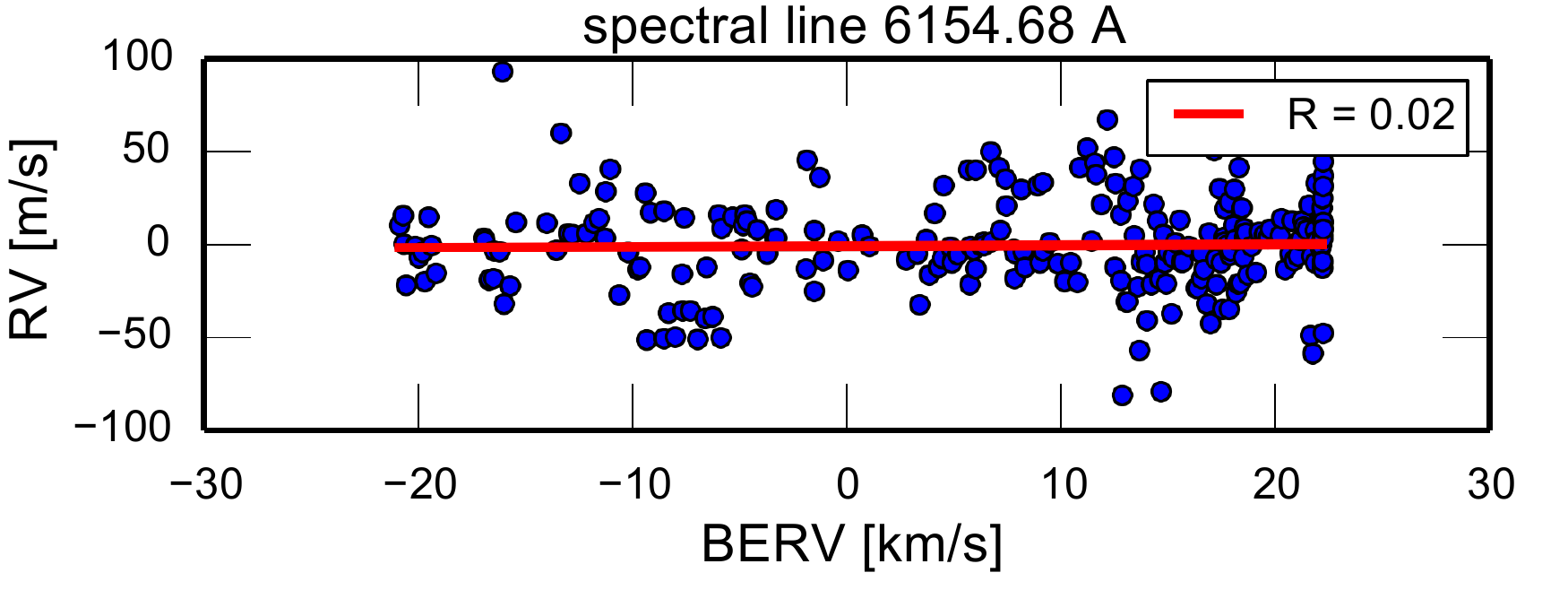}
\includegraphics[width=8cm]{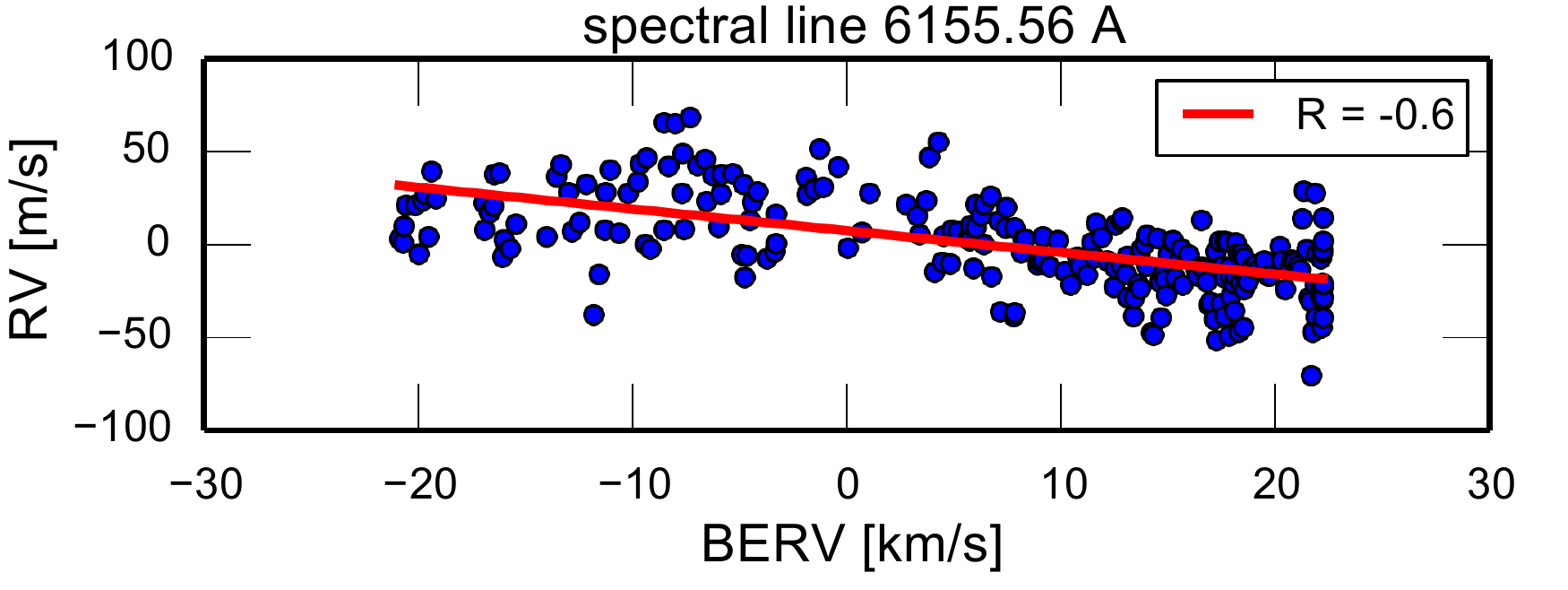}
\includegraphics[width=8cm]{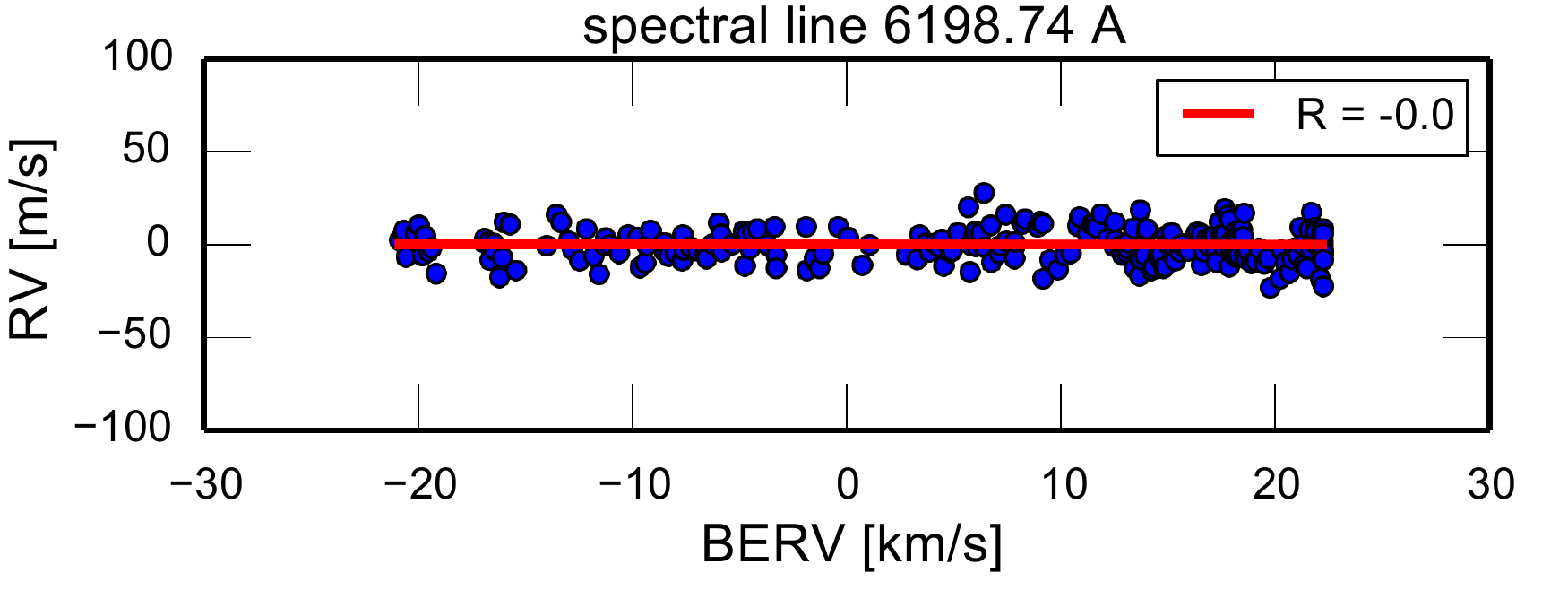}
\includegraphics[width=8cm]{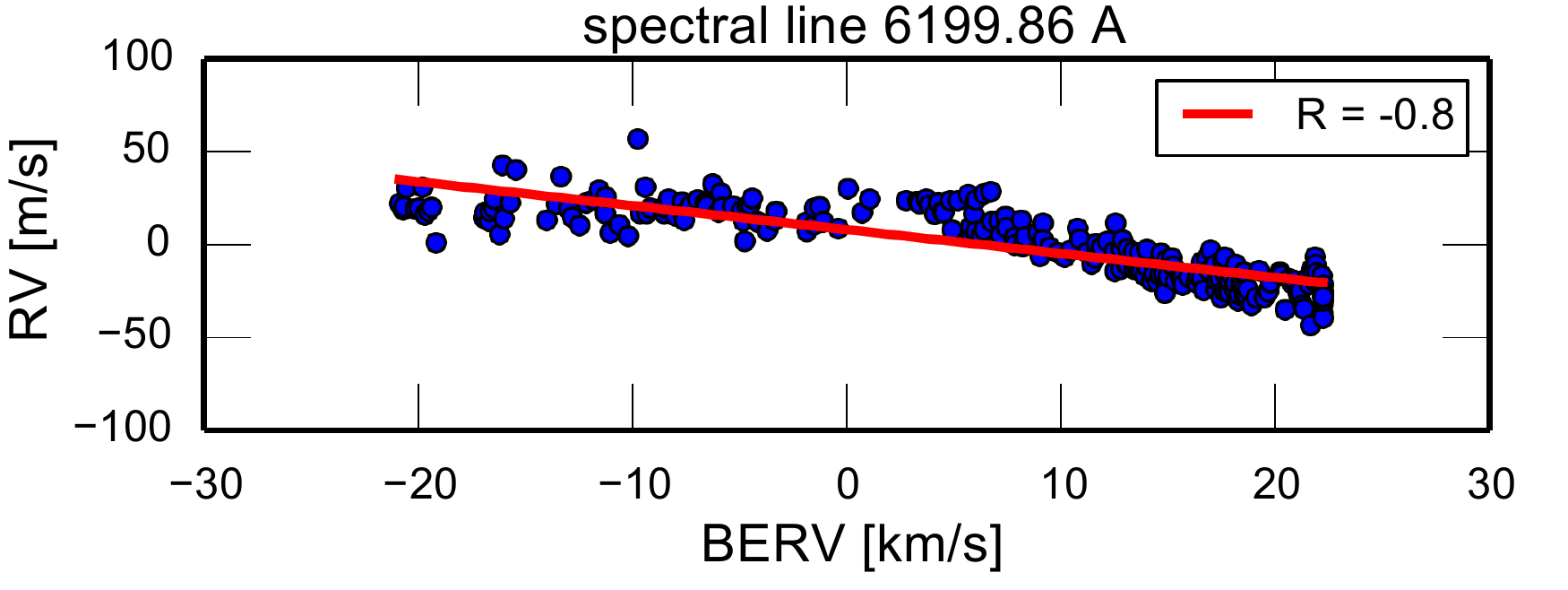}
\caption{Correlations between the RV derived from different spectral lines present in the HARPS spectral order 61 and the BERV for HD128621. The red lines represent a linear fit to the data, and the corresponding correlation coefficient is reported in the legend. A strong anti-correlation is observed for spectral lines at 6155.56 and 6199.86 \AA\,}
\label{fig:3-2}
\end{center}
\end{figure*}

To go a step further and detect which lines in spectral order 61 are responsible for the yearly signal, we fitted each individual line stronger than 20\% in depth with a Gaussian and selected the center of the Gaussian as the average wavelength of the line. This procedure is performed on all the stellar spectra to get the average wavelength position of each line as a function of time. The variation in average wavelength position as a function of time is then translated into a RV variation. Figure \ref{fig:3-1} shows the RV variation of four spectral lines as a function of time. The top two lines are very close in wavelength, {\bf however one of them shows a strong one-year signal, while the other does not}. The same behavior is observed for the two spectral lines at the bottom of Figure \ref{fig:3-1}. The two lines not affected by the yearly signal show no correlation between RV and BERV (see Figure. \ref{fig:3-2}), while the two other lines show a strong anti-correlation. This anti-correlation is similar to the one observed when using the RV derived with all the spectral lines in the spectrum (see Figure \ref{fig:2-2}), {\bf and therefore those two lines are examples of lines contributing to the detected yearly signal.}

\subsection{Origin of the CCD imperfections deforming spectral lines}  \label{sect:3-2}

The HARPS CCD is composed of a mosaic of two 2x4K EEV44-82 CCDs (\href{http://www.eso.org/sci/php/optdet/instruments/harps/FDR-Harps-detectors.pdf}{HARPS detector final design review\footnote{http://www.eso.org/sci/php/optdet/instruments/harps/FDR-Harps-detectors.pdf}}). Each CCD device is {\bf subdivided in 1024x512 blocks of pixels} (1024 in the cross-dispersion direction and 512 in the spectral direction) with a pixel size of 15\,$\mu$m within each block. The block boundaries show sometimes a 1\% quantum efficiency (QE) variation over 1 row or column due to photolithography stepper mismatches. This variation in QE is induced by a variation in pixel size at the area boundaries, due to difficulties to perform the stitching of all the blocks together. These boundaries will be called block stitchings hereafter.

To get a wavelength solution for each pixel illuminated by both spectrograph fibers, a Thorium-Argon lamp is used to illuminate each fiber. The emission lines in the Thorium-Argon lamp create a finger print on the CCD, and by knowing precisely the wavelength of each individual line, it is possible to derive a wavelength for all the pixels in each spectral order. Because Thorium-Argon emission lines are sparse on the detector, a {\bf polynomial fit is performed on each order} to get a wavelength for all pixels. {\bf The wavelength solution is correct if all pixels have the same size, which is not the case for the boundaries between the 1024x512 blocks constituting the CCD. This polynomial fit therefore introduces an error in wavelength solution at those boundaries}. This effect has been shown by \citet{Wilken-2010} (see Figure 4 of their paper), where the authors compare wavelength solutions obtained with a laser frequency comb and a Thorium-Argon lamp. Every 512 pixels in the spectral direction, a step as high as 40 \ms (1/20th of a pixel on HARPS) is observed when comparing the two wavelength solutions. These differences can be explained by the {\bf larger number of regular emission peaks} present in the laser frequency comb that can resolve the difference in pixel size at the block stitchings, which the spatially sparse Thorium-Argon emission lines can not do.

\begin{figure*}
\begin{center}
\includegraphics[width=8cm]{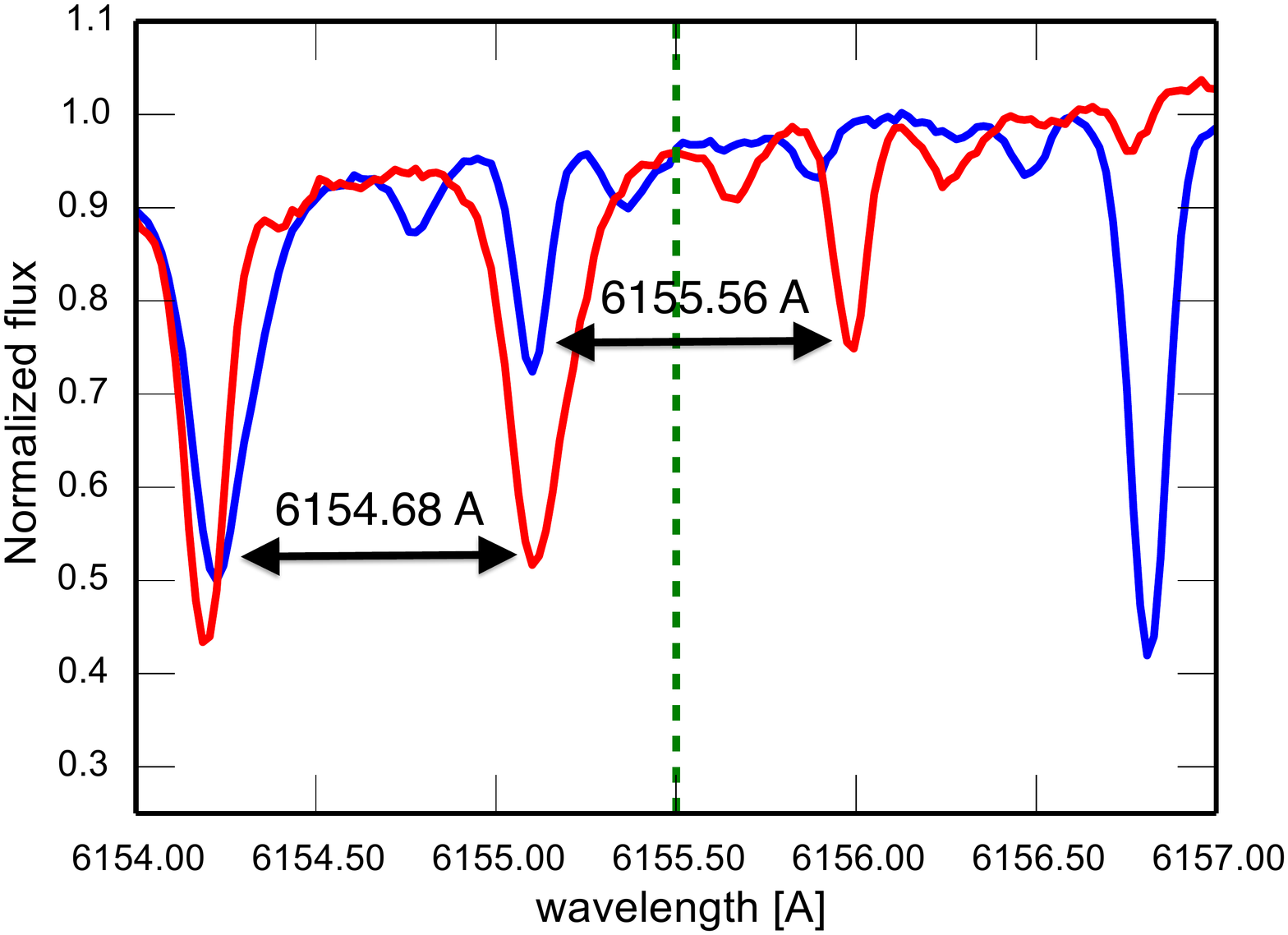}
\includegraphics[width=8cm]{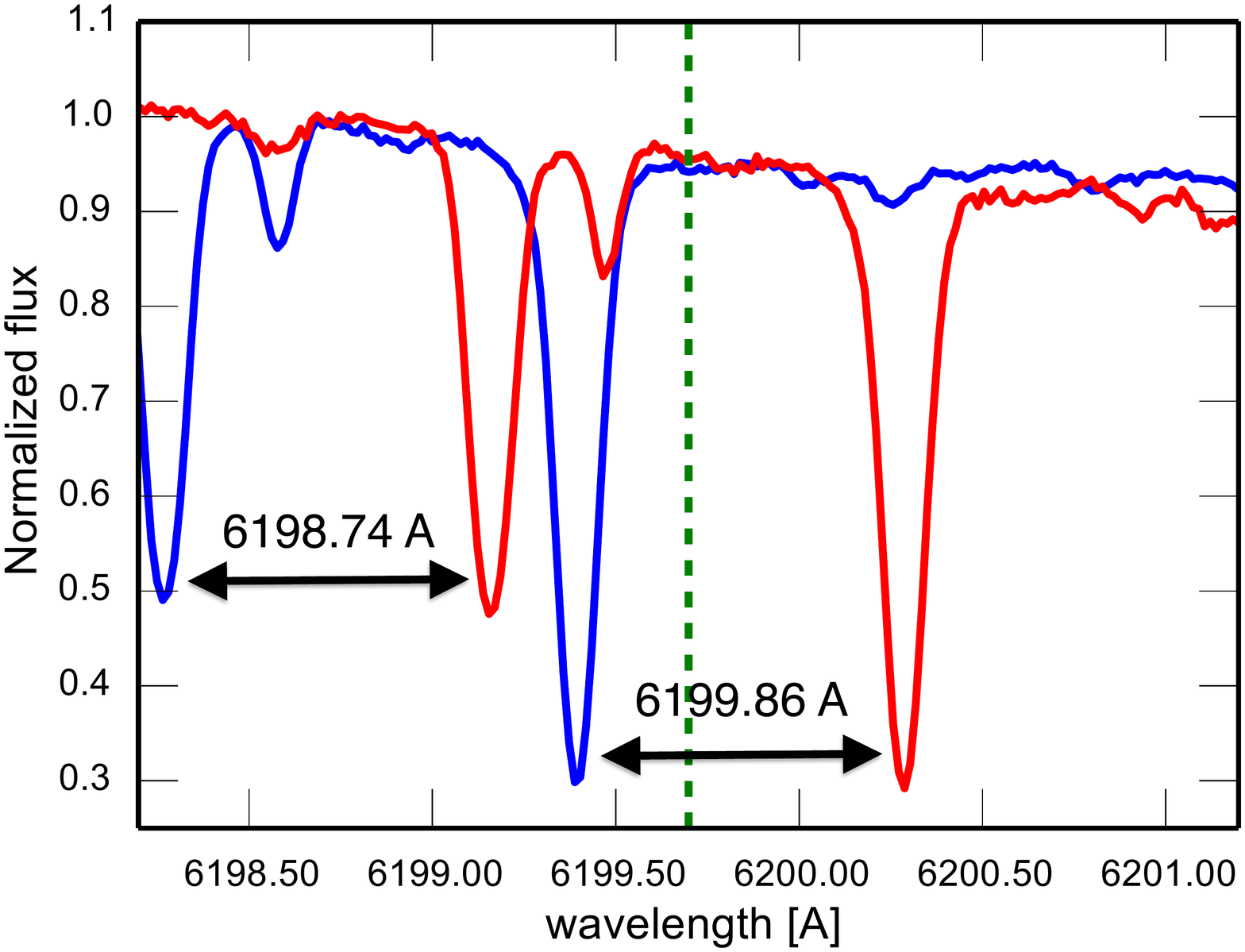}
\caption{Zoom on small spectral regions of spectral order 61. The blue spectrum corresponds to an observation of HD128621 when the BERV was maximum and the red spectrum when the BERV was minimum. The green dashed vertical lines corresponds to the position of the block stitchings. Spectral lines at 6155.56 and 6199.86 \AA\, showing the yearly effect cross these block stitchings, while the other lines, not affected, do not.}
\label{fig:3-3}
\end{center}
\end{figure*}

In Figure \ref{fig:3-3}, we show the part of the spectrum around both sets of spectral lines studied in the preceding section (6154.68 and 6155.56 \AA\,, and 6198.74 and 6199.86 \AA\,). The two different spectra in blue and red correspond to observations done at the extremes of the BERV, highlighting the shift of the spectrum on the CCD. As we can see, the spectral lines at 6155.56 and 6199.86 \AA\, affected by the yearly signal cross the block stitchings represented by the green vertical dashed line when Earth orbits the Sun. The two other lines do not cross those boundaries and therefore do not show the one-year variation.

\section{Correcting for the one-year signal} \label{sect:4}

In the preceding section, we demonstrated that the boundaries between the blocks constituting the CCD, so-called block stitchings, are responsible for the yearly signal detected in HD128621, HD1461, HD154088 and HD31527. A simple solution to get rid of this yearly signal is to remove the spectral lines passing over these boundaries from the correlation mask used to derive the RVs. Because each star has its own gamma velocity, the position of the spectral lines on the CCD will differ from one star to the other, and therefore a different correlation mask has to be generated for each star. 

To select the spectral lines at wavelength $\lambda_i$ that cross the block stitching when Earth is revolving around the Sun, the first step is to shift their wavelength to the stellar gamma velocity $\gamma_{\star}$. The new wavelength in the stellar rest frame $\lambda_{\star,i}$ can be computed using the Doppler effect formalism $\lambda_{\star,i}= \lambda_i \left(\sqrt{\frac{1+\gamma_{\star}/c}{1-\gamma_{\star}/c}} - 1\right)$. The second step is to identify the wavelength $\lambda_{BS,j}$ of the block stitchings, which can be done using the wavelength solution and keeping in mind that the block stitchings appear every 512 pixels in the spectral direction. Finally, all the spectral lines in the correlation mask that have a wavelength in the stellar rest frame obeying to the relation
\begin{eqnarray}
\lambda_{BS,j} - min_{BERV} - W \le \lambda_{\star,i} \le \lambda_{BS,j} + max_{BERV} + W
\end{eqnarray}
should be excluded. Here $W$ corresponds to the width of a spectral line. We fixed this value to 10\kms\, to exclude the wings of each spectral line \footnote{The average full width at half maximum of a line is a few \kms for a solar like star rotating slowly \citep[][]{Gray-2008}}. 

\begin{figure*}
\begin{center}
\includegraphics[width=16cm]{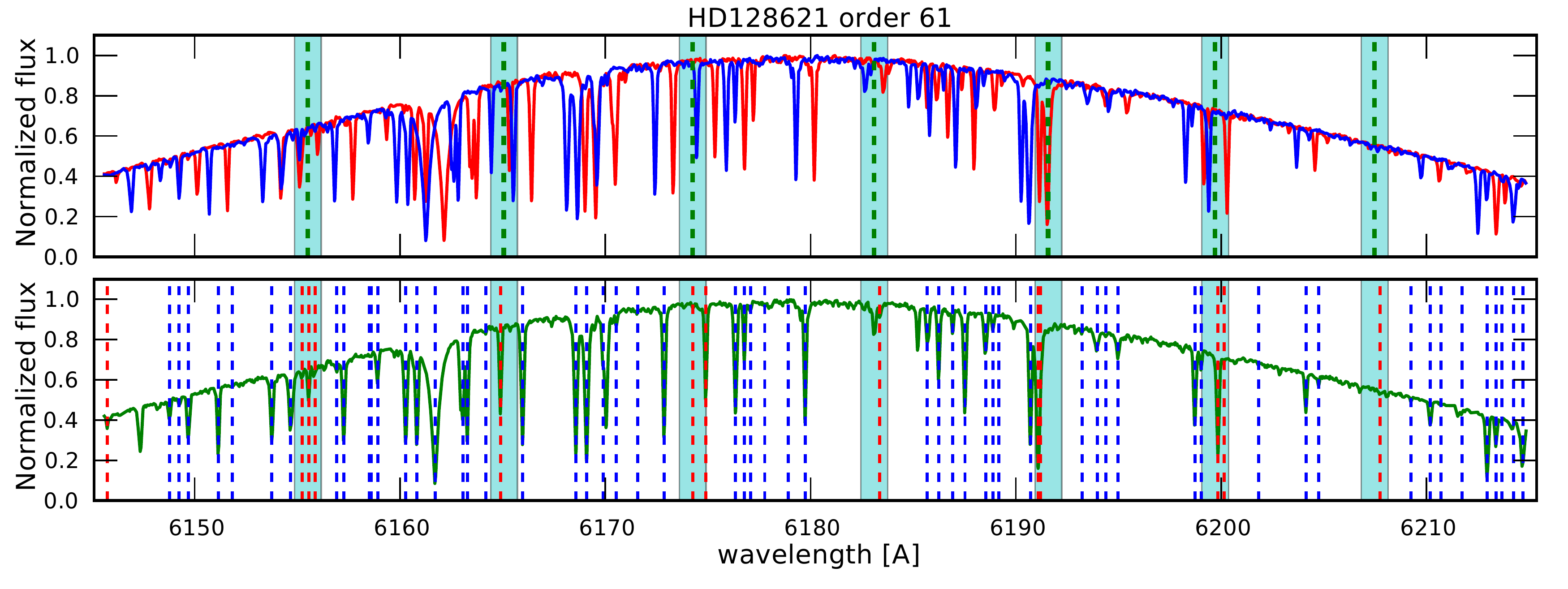}
\caption{\emph{Top: }The blue spectrum corresponds to spectral order 61 of an observation of HD128621 when the BERV was maximum and the red spectrum, spectral order 61 when the BERV was minimum. The green dashed vertical lines correspond to the position of the block stitchings and the blue regions to the zone where spectral lines should be excluded. \emph{Bottom: } Spectrum corresponding to order 61 of an observation of HD128621 when the BERV was close to zero. The dashed vertical lines are the spectral lines present in the correlation mask, shifted to the stellar gamma velocity. The spectral line falling inside the excluding regions are highlighted in red and should be excluded from the correlation mask when deriving RVs to prevent the one-year block stitching effects.}
\label{fig:4-0}
\end{center}
\end{figure*}

In Figure \ref{fig:4-0}, we illustrate the process of excluding the spectral lines from the correlation mask taking the example of spectral order 61. The top panel shows two spectra of HD128621 taken at the extremes of the BERV. Each block stitching is localized in the wavelength domain, and the blue region across each block stitching correspond to the region where spectral lines should be excluded. The bottom panel shows a spectrum of HD128621 taken when the BERV was close to zero. All the spectral lines appearing in the correlation mask and shifted to the stellar gamma velocity are shown, and the ones falling close to the block stitching are excluded (red lines). Note that on the blue side of the spectral order, a line is exclude while not falling on a block stitching. There is some overlap between the spectral orders, and this line is rejected in spectral order 60.

\begin{figure*}
\begin{center}
\includegraphics[width=8cm]{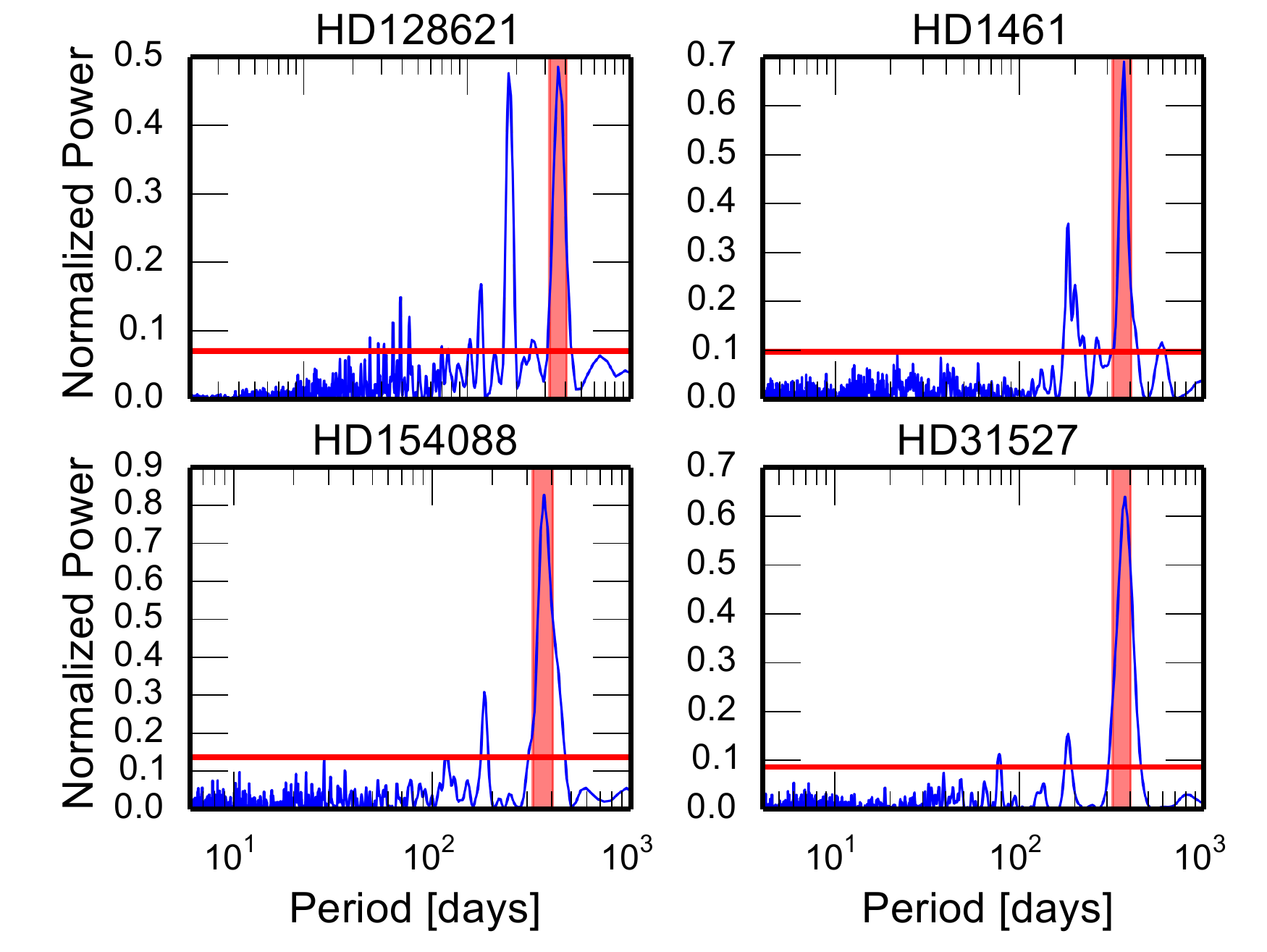}
\includegraphics[width=8cm]{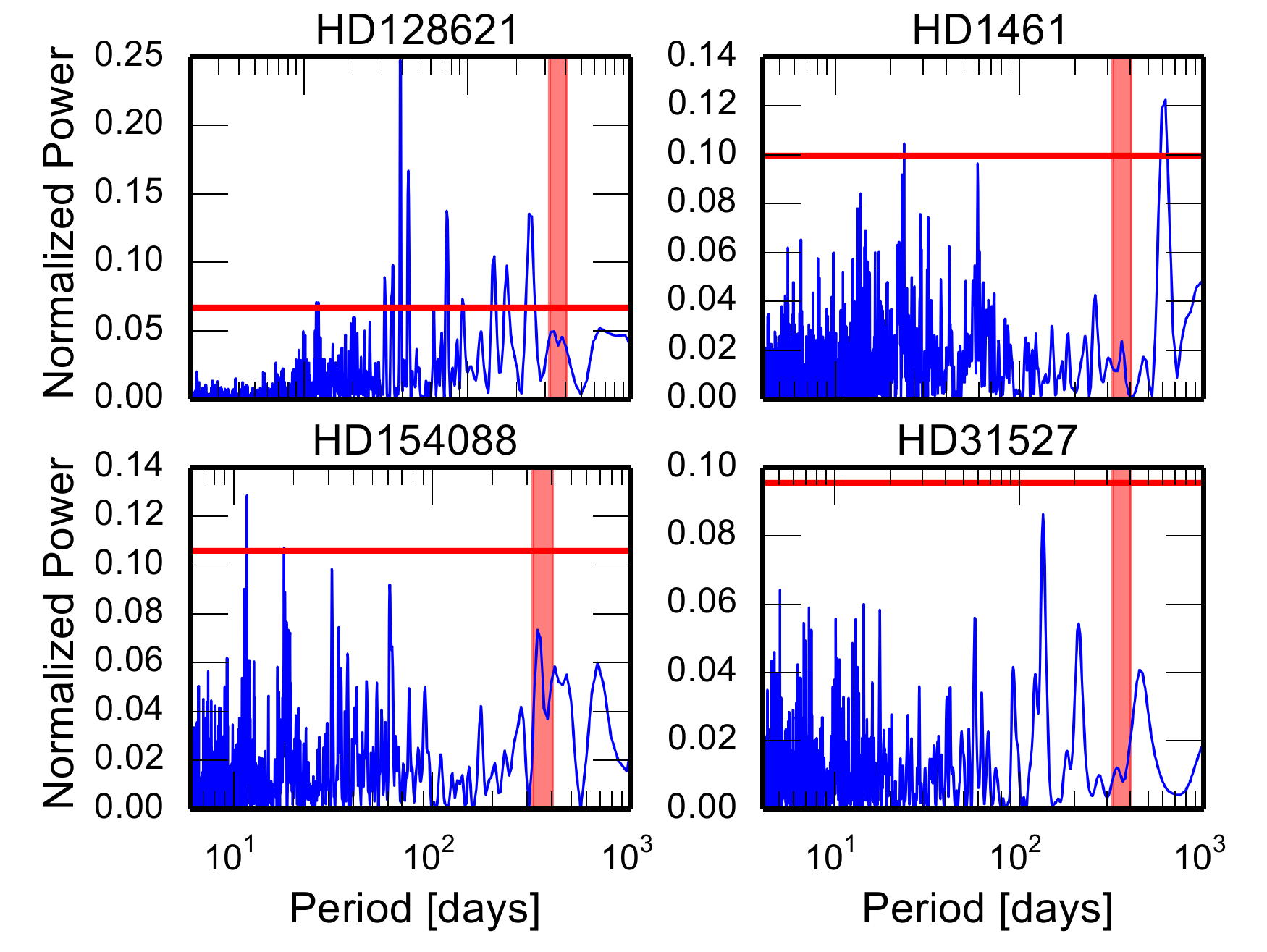}
\caption{\emph{Left: }GLS periodograms of the RV residuals of our four stars after removing their planetary signals and their magnetic cycle effect (see text in Sec \ref{sect:2} for more details about the fitted models). The RVs are derived using the correlation masks with only the spectral lines crossing the block stitchings. The periods close to one year are highlighted by the red regions, and are significantly above the 10\% false alarm probability level (red horizontal lines). \emph{Right: }Same plots but for the correlation masks cleaned from spectral lines crossing the block stitchings.}
\label{fig:4-1}
\end{center}
\end{figure*}

For each star, two different correlation masks are generated. One with only the lines falling close to the block stitchings, and another one without those lines. The RVs measured with both correlation masks can be seen in Figure \ref{fig:4-1}. It is clear that the spectral lines crossing the block stitching are the origin of the one-year signal (see left panel). Once those spectral lines are removed from the correlation mask, the yearly signal disappears (see right panel). Note that this Figure should be compared with Figure \ref{fig:2-0}.

\section{Conclusion}  \label{sect:5}

We have detected a one year signal in the data of several stars intensively observed with HARPS during several years. This signal is induced by spectral lines crossing the block stitchings of the HARPS CCD when the stellar spectrum is shifted back and forth on the detector due to the Earth's orbit around the Sun. At the position of each block stitching, the wavelength solution is not correct because the sparse emission spectrum of the Thorium-Argon lamp cannot resolve the different pixel size. Spectral lines crossing block stitchings therefore undergo a deformation when pixel position is translated into wavelength using the wavelength solution, which induces a RV variation. Removing those spectral lines from the correlation mask used to derive the RV with the cross-correlation technique allows to get rid of this yearly signal. Note that a different correlation mask for each star has to be generated due to a different stellar gamma velocity, and therefore a different position of the spectral lines on the HARPS CCD.

The amplitude of the one-year signal is different for each star, because of different BERV and different spectral lines crossing the block stitchings of the CCD (see Figure \ref{fig:2-2}). If the amplitude of the BERV is only a few \kms (maximum is 30\kms), then the stellar spectra will only be shifted on the CCD by a few pixels, which will not deform enough the spectral lines crossing the block stitchings and therefore not induce a significant signal at one year. In addition, if only weak spectral lines, that only have a small weight in the final RV computation as described in \citet{Pepe-2003}, cross block stitchings, the induced  RV variation will also be negligible. {\bf The combination of these two effects explains why some stars are strongly affected by the one-year signal, while others are effected to a lesser degree.}

The number of spectral lines rejected in the correlation mask is of the order of 20\%. This can therefore be a problem when observing faint targets, for which the signal-to-noise-ratio of the spectra is low. A possible way of removing the yearly signal without rejecting spectral lines would be to include as free parameters the pixel sizes at all block stitching, and thus derive a corrected wavelength solution (Coffinet et al. 2015, in prep.).


{\bf For calibration sources such as Fabry-Perot etalons or laser frequency-combs} \citep[][]{Langellier-2014,Wilken-2012} that have a higher density of emission lines on the detector, it is possible to detect the CCD block stitchings and therefore directly obtain a correct wavelength solution. In addition to a better RV precision, these calibration sources should produce data that are not affected by the one-year signal described in this paper.

This yearly signal is extremely problematic when looking for long-period small-amplitude planets as it will inject signal at 365.25 days and the first corresponding harmonics, generally 182.63 and 121.75 days. It is therefore mandatory to correct this yearly signal from the HARPS RV data if such planet want to be detected. {\bf For a user detecting a spurious signal at one year in HARPS RV data, two different methods can be used to suppress this signal. Using the pipeline, it is possible to recalculate a new correlation mask based on the technique described in Section 4 and reprocess the CCF routine. Otherwise fitting a sinusoidal signal with a one-year period is precise enough to a few dozens of \cms. A careful check of the anti-correlation between the fitted signal and the BERV should however be done, to prevent removing one-year planetary signals that could exist in the data.}

\acknowledgments
{\bf We thanks the referee for his valuable comments and suggestions that improved the first version of the paper. X.D. thanks NASA's Transiting Exoplanet Survey Satellite (TESS) mission for partial support via subaward 5710003554 from MIT to SAO.} We are grateful to all technical and scientific collaborators of the HARPS Consortium, ESO Headquarters and ESO La Silla who have contributed with their extraordinary passion and valuable work to the success of the HARPS project.

\bibliographystyle{apj}
\bibliography{dumusque_bibliography}

\begin{thebibliography}{}
\expandafter\ifx\csname natexlab\endcsname\relax\def\natexlab#1{#1}\fi

\bibitem[{{Arentoft} {et~al.}(2008){Arentoft}, {Kjeldsen}, {Bedding}, {Bazot},
  {Christensen-Dalsgaard}, {Dall}, {Karoff}, {Carrier}, {Eggenberger},
  {Sosnowska}, {Wittenmyer}, {Endl}, {Metcalfe}, {Hekker}, {Reffert}, {Butler},
  {Bruntt}, {Kiss}, {O'Toole}, {Kambe}, {Ando}, {Izumiura}, {Sato}, {Hartmann},
  {Hatzes}, {Bouchy}, {Mosser}, {Appourchaux}, {Barban}, {Berthomieu},
  {Garcia}, {Michel}, {Provost}, {Turck-Chi{\`e}ze}, {Marti{\'c}}, {Lebrun},
  {Schmitt}, {Bertaux}, {Bonanno}, {Benatti}, {Claudi}, {Cosentino}, {Leccia},
  {Frandsen}, {Brogaard}, {Glowienka}, {Grundahl}, \&
  {Stempels}}]{Arentoft-2008}
{Arentoft}, T., {Kjeldsen}, H., {Bedding}, T.~R., {et~al.} 2008, \apj, 687,
  1180

\bibitem[{{Boisse} {et~al.}(2012){Boisse}, {Bonfils}, \&
  {Santos}}]{Boisse-2012b}
{Boisse}, I., {Bonfils}, X., \& {Santos}, N.~C. 2012, \aap, 545, 109

\bibitem[{{Cosentino} {et~al.}(2012){Cosentino}, {Lovis}, {Pepe}, {Collier
  Cameron}, {Latham}, {Molinari}, {Udry}, {Bezawada}, {Black}, {Born},
  {Buchschacher}, {Charbonneau}, {Figueira}, {Fleury}, {Galli}, {Gallie},
  {Gao}, {Ghedina}, {Gonzalez}, {Gonzalez}, {Guerra}, {Henry}, {Horne},
  {Hughes}, {Kelly}, {Lodi}, {Lunney}, {Maire}, {Mayor}, {Micela}, {Ordway},
  {Peacock}, {Phillips}, {Piotto}, {Pollacco}, {Queloz}, {Rice}, {Riverol},
  {Riverol}, {San Juan}, {Sasselov}, {Segransan}, {Sozzetti}, {Sosnowska},
  {Stobie}, {Szentgyorgyi}, {Vick}, \& {Weber}}]{Cosentino-2012}
{Cosentino}, R., {Lovis}, C., {Pepe}, F., {et~al.} 2012, in Society of
  Photo-Optical Instrumentation Engineers (SPIE) Conference Series, Vol. 8446,
  Society of Photo-Optical Instrumentation Engineers (SPIE) Conference Series

\bibitem[{{Dumusque} {et~al.}(2011{\natexlab{a}}){Dumusque}, {Udry}, {Lovis},
  {Santos}, \& {Monteiro}}]{Dumusque-2011a}
{Dumusque}, X., {Udry}, S., {Lovis}, C., {Santos}, N.~C., \& {Monteiro},
  M.~J.~P.~F.~G. 2011{\natexlab{a}}, \aap, 525, A140

\bibitem[{{Dumusque} {et~al.}(2011{\natexlab{b}}){Dumusque}, {Lovis},
  {S{\'e}gransan}, {Mayor}, {Udry}, {Benz}, {Bouchy}, {Lo Curto}, {Mordasini},
  {Pepe}, {Queloz}, {Santos}, \& {Naef}}]{Dumusque-2011c}
{Dumusque}, X., {Lovis}, C., {S{\'e}gransan}, D., {et~al.} 2011{\natexlab{b}},
  \aap, 535, A55

\bibitem[{{Dumusque} {et~al.}(2012){Dumusque}, {Pepe}, {Lovis}, {Segransan},
  {Sahlmann}, {Benz}, {Bouchy}, {Mayor}, {Queloz}, {Santos}, \&
  {Udry}}]{Dumusque-2012}
{Dumusque}, X., {Pepe}, F., {Lovis}, C., {et~al.} 2012, \nat, 491, 207

\bibitem[{{Dumusque} {et~al.}(2014){Dumusque}, {Bonomo}, {Haywood},
  {Malavolta}, {S{\'e}gransan}, {Buchhave}, {Collier Cameron}, {Latham},
  {Molinari}, {Pepe}, {Udry}, {Charbonneau}, {Cosentino}, {Dressing},
  {Figueira}, {Fiorenzano}, {Gettel}, {Harutyunyan}, {Horne}, {Lopez-Morales},
  {Lovis}, {Mayor}, {Micela}, {Motalebi}, {Nascimbeni}, {Phillips}, {Piotto},
  {Pollacco}, {Queloz}, {Rice}, {Sasselov}, {Sozzetti}, {Szentgyorgyi}, \&
  {Watson}}]{Dumusque-2014a}
{Dumusque}, X., {Bonomo}, A.~S., {Haywood}, R.~D., {et~al.} 2014, \apj, 789,
  154

\bibitem[{{Gray}(2008)}]{Gray-2008}
{Gray}, D.~F. 2008, {The Observation and Analysis of Stellar Photospheres}, ed.
  U.~C. U.~P. Cambridge

\bibitem[{{Kjeldsen} {et~al.}(2005){Kjeldsen}, {Bedding}, {Butler},
  {Christensen-Dalsgaard}, {Kiss}, {McCarthy}, {Marcy}, {Tinney}, \&
  {Wright}}]{Kjeldsen-2005}
{Kjeldsen}, H., {Bedding}, T.~R., {Butler}, R.~P., {et~al.} 2005, \apj, 635,
  1281

\bibitem[{{Kopeikin} \& {Ozernoy}(1999)}]{Kopeikin-1999}
{Kopeikin}, S.~M., \& {Ozernoy}, L.~M. 1999, \apj, 523, 771

\bibitem[{{Langellier} {et~al.}(2014){Langellier}, {Li}, {Glenday}, {Chang},
  {Chen}, {Lim}, {Furesz}, {K{\"a}rtner}, {Phillips}, {Sasselov},
  {Szentgyorgyi}, \& {Walsworth}}]{Langellier-2014}
{Langellier}, N., {Li}, C.-H., {Glenday}, A.~G., {et~al.} 2014, in Society of
  Photo-Optical Instrumentation Engineers (SPIE) Conference Series, Vol. 9147,
  Society of Photo-Optical Instrumentation Engineers (SPIE) Conference Series,
  8

\bibitem[{{Lindegren} \& {Dravins}(2003)}]{Lindegren-2003}
{Lindegren}, L., \& {Dravins}, D. 2003, \aap, 401, 1185

\bibitem[{{Mayor} {et~al.}(2003){Mayor}, {Pepe}, {Queloz}, {Bouchy},
  {Rupprecht}, {Lo Curto}, {Avila}, {Benz}, {Bertaux}, {Bonfils}, {Dall},
  {Dekker}, {Delabre}, {Eckert}, {Fleury}, {Gilliotte}, {Gojak}, {Guzman},
  {Kohler}, {Lizon}, {Longinotti}, {Lovis}, {Megevand}, {Pasquini}, {Reyes},
  {Sivan}, {Sosnowska}, {Soto}, {Udry}, {van Kesteren}, {Weber}, \&
  {Weilenmann}}]{Mayor-2003}
{Mayor}, M., {Pepe}, F., {Queloz}, D., {et~al.} 2003, The Messenger, 114, 20

\bibitem[{{Mayor} {et~al.}(2011){Mayor}, {Marmier}, {Lovis}, {Udry},
  {S{\'e}gransan}, {Pepe}, {Benz}, {Bertaux}, {Bouchy}, {Dumusque}, {Lo Curto},
  {Mordasini}, {Queloz}, \& {Santos}}]{Mayor-2011}
{Mayor}, M., {Marmier}, M., {Lovis}, C., {et~al.} 2011, ArXiv e-prints,
  arXiv:1109.2497

\bibitem[{{Meunier} {et~al.}(2010){Meunier}, {Desort}, \&
  {Lagrange}}]{Meunier-2010a}
{Meunier}, N., {Desort}, M., \& {Lagrange}, A.-M. 2010, \aap, 512, A39

\bibitem[{{Meunier} \& {Lagrange}(2013)}]{Meunier-2013}
{Meunier}, N., \& {Lagrange}, A.-M. 2013, \aap, 551, A101

\bibitem[{{Pepe} {et~al.}(2002){Pepe}, {Mayor}, {Rupprecht}, {Avila},
  {Ballester}, {Beckers}, {Benz}, {Bertaux}, {Bouchy}, {Buzzoni}, {Cavadore},
  {Deiries}, {Dekker}, {Delabre}, {D'Odorico}, {Eckert}, {Fischer}, {Fleury},
  {George}, {Gilliotte}, {Gojak}, {Guzman}, {Koch}, {Kohler}, {Kotzlowski},
  {Lacroix}, {Le Merrer}, {Lizon}, {Lo Curto}, {Longinotti}, {Megevand},
  {Pasquini}, {Petitpas}, {Pichard}, {Queloz}, {Reyes}, {Richaud}, {Sivan},
  {Sosnowska}, {Soto}, {Udry}, {Ureta}, {van Kesteren}, {Weber}, {Weilenmann},
  {Wicenec}, {Wieland}, {Christensen-Dalsgaard}, {Dravins}, {Hatzes},
  {K{\"u}rster}, {Paresce}, \& {Penny}}]{Pepe-2002}
{Pepe}, F., {Mayor}, M., {Rupprecht}, G., {et~al.} 2002, The Messenger, 110, 9

\bibitem[{{Pepe} {et~al.}(2003){Pepe}, {Rupprecht}, {Avila}, {Balestra},
  {Bouchy}, {Cavadore}, {Eckert}, {Fleury}, {Gillotte}, {Gojak}, {Guzman},
  {Kohler}, {Lizon}, {Mayor}, {Megevand}, {Queloz}, {Sosnowska}, {Udry}, \&
  {Weilenmann}}]{Pepe-2003}
{Pepe}, F., {Rupprecht}, G., {Avila}, G., {et~al.} 2003, in Society of
  Photo-Optical Instrumentation Engineers (SPIE) Conference Series, Vol. 4841,
  Society of Photo-Optical Instrumentation Engineers (SPIE) Conference Series,
  ed. M.~{Iye} \& A.~F.~M. {Moorwood}, 1045--1056

\bibitem[{{Pepe} {et~al.}(2011){Pepe}, {Lovis}, {S{\'e}gransan}, {Benz},
  {Bouchy}, {Dumusque}, {Mayor}, {Queloz}, {Santos}, \& {Udry}}]{Pepe-2011}
{Pepe}, F., {Lovis}, C., {S{\'e}gransan}, D., {et~al.} 2011, \aap, 534, A58

\bibitem[{{Saar} \& {Donahue}(1997)}]{Saar-1997b}
{Saar}, S.~H., \& {Donahue}, R.~A. 1997, \apj, 485, 319

\bibitem[{{Wildi} {et~al.}(2011){Wildi}, {Pepe}, {Chazelas}, {Lo Curto}, \&
  {Lovis}}]{Wildi-2011}
{Wildi}, F., {Pepe}, F., {Chazelas}, B., {Lo Curto}, G., \& {Lovis}, C. 2011,
  in Society of Photo-Optical Instrumentation Engineers (SPIE) Conference
  Series, Vol. 8151, Society of Photo-Optical Instrumentation Engineers (SPIE)
  Conference Series, 1

\bibitem[{{Wilken} {et~al.}(2010){Wilken}, {Lovis}, {Manescau}, {Steinmetz},
  {Pasquini}, {Lo Curto}, {H{\"a}nsch}, {Holzwarth}, \& {Udem}}]{Wilken-2010}
{Wilken}, T., {Lovis}, C., {Manescau}, A., {et~al.} 2010, \mnras, 405, L16

\bibitem[{{Wilken} {et~al.}(2012){Wilken}, {Curto}, {Probst}, {Steinmetz},
  {Manescau}, {Pasquini}, {Gonz{\'a}lez Hern{\'a}ndez}, {Rebolo}, {H{\"a}nsch},
  {Udem}, \& {Holzwarth}}]{Wilken-2012}
{Wilken}, T., {Curto}, G.~L., {Probst}, R.~A., {et~al.} 2012, \nat, 485, 611

\bibitem[{{Wright} \& {Eastman}(2014)}]{Wright-2014}
{Wright}, J.~T., \& {Eastman}, J.~D. 2014, \pasp, 126, 838

\bibitem[{{Zechmeister} \& {K{\"u}rster}(2009)}]{Zechmeister-2009}
{Zechmeister}, M., \& {K{\"u}rster}, M. 2009, \aap, 496, 577

\end{thebibliography}

\end{document}